%% file: main.tex
\documentclass[journal]{IEEEtran}

\usepackage{xcolor} 
\usepackage{url}
\usepackage{paralist} 
\usepackage{tcolorbox} 
\usepackage{tabularray}
\usepackage{graphicx} 
\usepackage{multirow}
\usepackage{hyperref} 
\usepackage{makecell} 
\usepackage{xspace}
\usepackage{balance}
\usepackage{caption} 
\input{solidity/solidity-highlighting}

\usepackage{amssymb}
\usepackage{pifont}
\usepackage{listings}

\usepackage{xcolor}
\definecolor{eclipseKeywords}{RGB}{127,0,85}
\usepackage[export]{adjustbox}
\colorlet{numb}{magenta!60!black}

\lstdefinelanguage{json}{
    basicstyle=\normalfont\ttfamily,
    numbers=left,
    numberstyle=\scriptsize,
    stepnumber=1,
    numbersep=5pt,
    tabsize=2, 
    showstringspaces=false,
    breaklines=true,
    frame=lines,
    string=[s]{"}{"},
    comment=[l]{:\ "},
    morecomment=[l]{:"},
    literate=
        *{0}{{{\color{numb}0}}}{1}
         {1}{{{\color{numb}1}}}{1}
         {2}{{{\color{numb}2}}}{1}
         {3}{{{\color{numb}3}}}{1}
         {4}{{{\color{numb}4}}}{1}
         {5}{{{\color{numb}5}}}{1}
         {6}{{{\color{numb}6}}}{1}
         {7}{{{\color{numb}7}}}{1}
         {8}{{{\color{numb}8}}}{1}
         {9}{{{\color{numb}9}}}{1}
}

\definecolor{darkgreen}{rgb}{0.0, 0.5, 0.0}
\newcommand{\cmark}{\textcolor{darkgreen}{\ding{51}}}%
\newcommand{\xmark}{\textcolor{red}{\ding{55}}}%
\newtcolorbox{answerbox}[1]{colback=gray!10!white,coltitle=black,
title={#1},fonttitle=\bfseries,attach title to upper,after title={:\ }}

\newcommand{\revision}[1]{\textcolor{black}{#1}}

\newenvironment{rblock}
  {\begingroup
   \color{black}
   \captionsetup{labelfont={color=black}, textfont={color=black}}%
  }
  {\endgroup}

\newcommand{\totaltools}{\textsc{19}\xspace}
\newcommand{\toolsavailable}{\textsc{15}\xspace}
\newcommand{\totalinstall}{\textsc{11}\xspace}
\newcommand{\totalreproducible}{\textsc{7}\xspace}
\newcommand{\totalexecutable}{\textsc{7}\xspace}

\newcommand{\totalcontracts}{\textsc{143}\xspace}

\newcommand{\rqone}{To what extent are research tools for APR smart contract available and executable?\xspace}
\newcommand{\rqtwo}{How do APR tools for smart contracts perform on a common benchmark?\xspace}
\newcommand{\rqthree}{To what extent do APR tools mitigate executable smart contract exploits?\xspace}
\newcommand{\rqfour}{What is the mitigation effectiveness per vulnerability type?\xspace}

\newcommand{\rqfive}{\revision{ For patches that mitigate an exploit, to what extent can they be considered correct?}\xspace}

\newcommand{\dataset}{SmartBugs-Curated\xspace}

\newcommand{\exploitdatasetname}{sb-heists\xspace}

\begin{document}

\title{Do Automated Fixes Truly Mitigate\\ Smart Contract Exploits?}
\author{Sofia Bobadilla$^1$, Monica Jin$^1$, Martin Monperrus\\KTH Royal Institute of Technology\\\texttt{\{sofbob,mjin,monperrus\}@kth.se}\vspace{-1cm}
 }


 \maketitle

\begin{abstract}
Automated Program Repair (APR) for smart contract security promises to automatically mitigate smart contract vulnerabilities responsible for billions in financial losses.
However, the true effectiveness of this research in addressing smart contract exploits remains uncharted territory.
This paper bridges this critical gap by introducing a novel and systematic experimental framework for evaluating exploit mitigation of program repair tools for smart contracts.

We qualitatively and quantitatively analyze 20 state-of-the-art APR tools using a dataset of 143 vulnerable smart contracts, for which we manually craft 91 executable exploits.
We are the very first to define and measure the essential ``exploit mitigation rate'', giving researchers and practitioners a real sense of effectiveness of cutting edge techniques.
Our findings reveal substantial disparities in the state of the art, with an exploit mitigation rate ranging from a low of 29\% to a high of 74\%.
Our study identifies systemic limitations, such as inconsistent functionality preservation, that must be addressed in future research on program repair for smart contracts.
\end{abstract}


%
\IEEEpeerreviewmaketitle

\def\thefootnote{1}\footnotetext{The authors are ordered alphabetically and contributed as follows. All authors defined the research questions and designed the empirical study. M.J. led the execution of the empirical study, including designing and implementing the experimental framework, debugging and reproducing the repair tools, implementing the exploits, and conducting the experiments. S.B. contributed to the reproduction of tools, the implementation of the exploits, and the review of the related literature. S.B. and M.J. both contributed to interpreting, analyzing the experimental results and performing the manual analysis. M.M. supervised the project and provided methodological guidance. All authors participated in writing and revising the manuscript.
}\def\thefootnote{\arabic{footnote}}
\section{Introduction}

\IEEEPARstart{T}{he} 
cost of deploying a vulnerable smart contract on-chain can reach millions of dollars in financial losses.
Hence, smart contract developers and auditors dedicate significant time and effort to ensuring that the contracts are free from vulnerabilities prior to deployment~\cite{vulnotexploit}.
There are multiple ways smart contracts can be attacked.
For example, a reentrancy problem is one of the ecosystem’s most notorious smart contract vulnerabilities and keeps being exploited \cite{2024surveydetectiontechniques}.

\revision{A key reason why smart contract vulnerabilities persist is the lack of effective tooling. Developers frequently cite insufficient automated support as a major obstacle to writing secure contracts~\cite{contractsdefect,challengesandopportunities}. One kind of tooling is Automated Program Repair (APR) for smart contracts \cite{survey-data-detection-repair-2023,qian2023empiricalreviewsmartcontract}, which promises to automatically patch vulnerabilities, offering a scalable way to secure contracts and reduce reliance on expensive manual audits.
}

\revision{However, despite this promise, APR for smart contracts remains underdeveloped compared to general purpose program repair~\cite{2023-learning-based-apr-survey,livingreview}. 
Most notably, there is no accepted sound methodology for evaluating whether an automated repair smart contract patch is effective~\cite{survey-data-detection-repair-2023,qian2023empiricalreviewsmartcontract}.
This methodological weakness has three critical consequences.
First, researchers cannot reliably measure progress in the field because the published results are based on incompatible evaluation setups.
Second, practitioners cannot confidently trust or adopt repair tools, as there is no solid evidence of their comparative effectiveness.
Third, the community lacks consolidated knowledge, which impedes not just research but also training, education, and ecosystem-wide improvements.
}

Our paper fills this crucial methodological gap, as we present a framework for sound empirical comparison among APR tools. The framework is built over one of the most thorough literature reviews of the field, and a novel experimental methodology to validate patches and vulnerability mitigation through actual exploits.
The core innovation is our focus on executable exploits as ground truth for patch validation.
Rather than relying on static analysis or manual inspection, we verify repair effectiveness by attempting to execute exploits against the patched contracts.
This approach provides concrete evidence of whether an automated fix to a vulnerable contract prevents the corresponding attack.

We reproduce 7 out of 20 state-of-the-art APR tools and apply our methodology in an extensive empirical study using a dataset of 143 vulnerable contracts.
We manually craft 91 executable exploits for these contracts.
Our evaluation reveals that effectiveness varies significantly from 29\% to 74\%, across different vulnerability types and repair strategies.

Our work sets the ground for further research on automated program repair for smart contracts by providing a rigorous theoretical and empirical comparison among APR tools.  
For researchers and practitioners alike, our evaluation framework and the publicly available exploit dataset provide an essential foundation for assessing and enhancing smart contract security.


To summarize, we make the following contributions:
\begin{itemize}
    \item A comprehensive literature review of automated program repair for smart contracts.
    \item A new methodology for measuring repair effectiveness through performing actual smart contract exploit mitigation, based on real, executable exploits.
    \item A complete set of experiments for our methodology based on \totalcontracts contracts, \totalreproducible APR tools, and 91 exploits. 
    \item \revision{A publicly available replication package: 1) set of exploits for the \dataset dataset, with a well-designed exploitation harness, \exploitdatasetname at \url{https://github.com/ASSERT-KTH/sb-heists}, 2) all generated patches and corresponding manual analysis results at \url{https://github.com/ASSERT-KTH/RepairComp/}.}
\end{itemize}

\revision{The rest of this paper is organized as follows.
Section~\ref{sec:background} presents the knowledge required to understand our study.
Section~\ref{sec:problem_space} surveys the problem space of program repair for smart contracts.
Section~\ref{sec:experimental_methodology} details our evaluation methodology.
Section~\ref{sec:experimental_results} presents our empirical results.
Section~\ref{sec:discussion} discusses implications.
Section~\ref{sec:validity} presents the threats to validity of our work.
Section~\ref{sec:related_work} covers related work, and Section~\ref{sec:conclusion} concludes.}

\begin{rblock}
\section{Background}\label{sec:background}
This section introduces the key concepts needed to understand our experimental setup and the implications of our findings.

\textbf{Smart Contracts.}
Smart Contracts are self-executing programs, deployed on blockchain networks, to enforce predefined rules without intermediaries. 
Contracts written in high-level smart contract languages like Solidity are compiled to bytecode for deployment and execution.

\textbf{Solidity.}
Solidity is Ethereum’s main smart contract programming language. 
While expressive and tailored for the EVM, misuse of constructs like fallback functions and low-level calls can lead to security vulnerabilities.
Although newer versions have introduced stricter checks and safer defaults, certain vulnerabilities are tightly linked to the language's design and semantics.

\textbf{Ethereum Virtual Machine.}
The Ethereum Virtual Machine (EVM) is a Turing-complete runtime environment that deterministically executes smart contract bytecode on the Ethereum blockchain.
It operates as a transaction-based state machine, where all state transitions, including smart contract deployment and execution, are triggered by transactions.

\textbf{Transactions.}
Transactions are initiated by an Externally Owned Account (EOA).
An EOA, controlled by a user through a private key, can initiate transactions to transfer Ether, deploy smart contracts, or invoke functions on existing contracts.
During execution, a called contract may interact with other contracts.
These interactions, known as internal transactions, occur within the same atomic execution context where either all succeed or all fail.

\textbf{Smart Contract Vulnerabilities.}
There is a wide range of smart contract vulnerabilities \cite{2024surveyethscsecurity-attackochdetection}.
A notable example is reentrancy, which arises when a contract calls an external contract before updating its internal state. 
Attackers can exploit this by invoking recursive calls through fallback functions, enabling malicious behavior such as fund draining. 
The 2016 DAO hack, which resulted in \$50 million in stolen Ether, exemplifies the severity of such flaws.
Other notable types of smart contract vulnerabilities include bad randomness \cite{bad_randomness}, numerical defects \cite{numericaldefects}, and price oracle manipulation \cite{priceoraclemanipulation}.


\end{rblock}

\section{The Problem Space of Program Repair for Smart Contracts}
\label{sec:problem_space}

\begin{figure}[ht!]
    \centering
    \includegraphics[width=\linewidth]{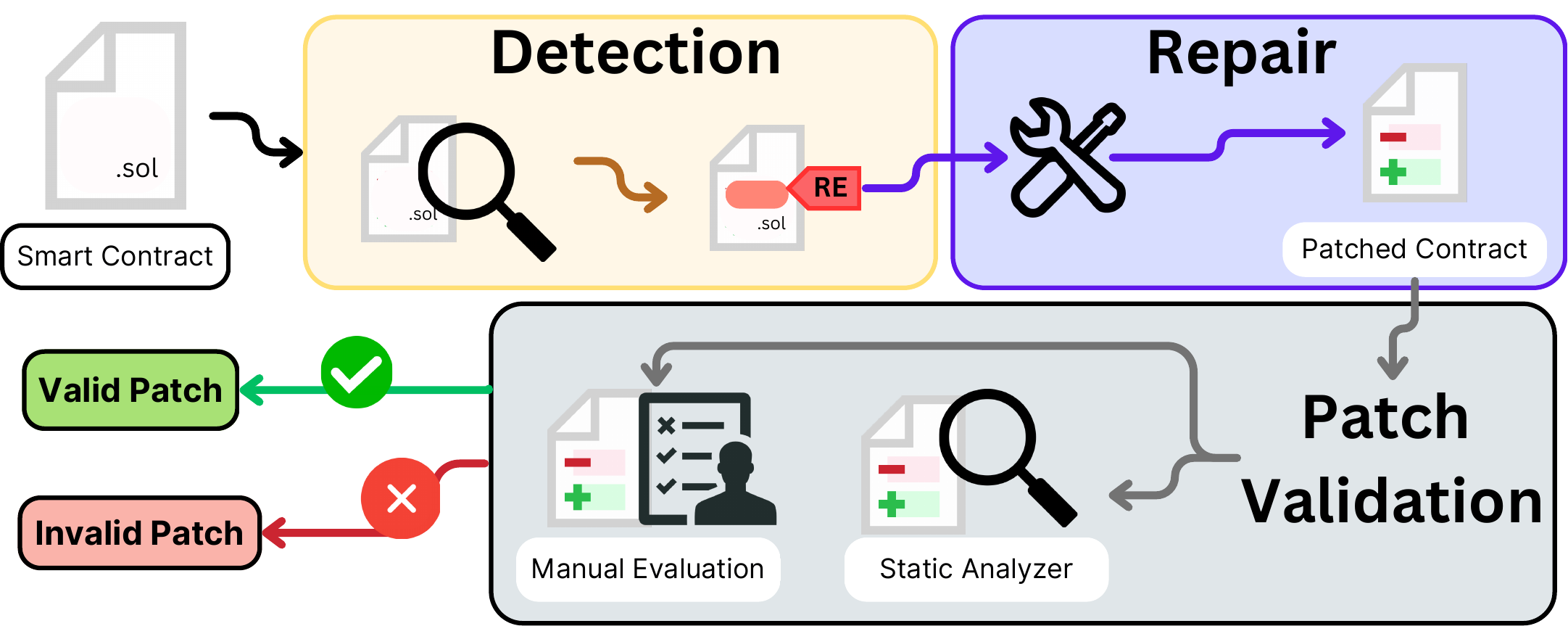}
    \caption{Overview of the general workflow of an automated program repair tool for smart contracts.}
    \label{fig:enter-label}
\end{figure}
\subsection{Overview}
The goal of Automated Program Repair (APR) tools for smart contracts is to generate a patch for a smart contract that is affected by a vulnerability. These tools have 3 key components:  \begin{inparaenum}
\item Detection: To repair a piece of code, the tool must know what needs to be fixed, using some existing techniques.
\item Repair: Once the tool has received the localization of the vulnerability to be repaired, it must apply changes to the original code.
\item Patch Validation: An APR tool must output a patch that solves a vulnerability. There are different ways for assessing the validity of a patch.
\end{inparaenum}

Our literature review describes the tools from the research landscape.
Table~\ref{tab:ToolDescription} consolidates our findings, highlighting the key components of each tool. 
Under “Repair-Level", we specify the patch type that each tool generates (see \autoref{subsec:level}). 
For “Detection", “Repair Strategy", and “Patch Validation" we follow the categories outlined in \ref{subsec:vulnerabilitydetection},\ref{subsec:repairstrategy} and \ref{subsec:patchvalidation}, respectively.

\subsection{Program Repair Tools for Smart Contracts}

\input{tables/Introduction/tools}

We now identify all notable program repair tools for smart contracts from the literature. Our systematic approach to select these tools is composed of \begin{inparaenum}[1)]
    \item Searching for the keywords \textit{program repair} and \textit{smart contracts} on Google Scholar , 
    \item Reading all collected papers,
    \item Performing citation analysis from their bibliographies,
    \item Assessing whether they are compatible with the Ethereum stack for the sake of comparative evaluation,
    \item Identifying tool names and links.
\end{inparaenum}

Through this rigorous investigation, we identified \totaltools potential APR tools for smart contracts. To the best of our knowledge, this is the most comprehensive list of smart contract APR tools at the time of writing, October 1st, 2024.

The first automated program repair tool was \textbf{SCRepair~\cite{SCRepair}}, which proposes a mutation-based approach to generate patches and validate them using a test suite. Also working with mutation repair, \textbf{DeFinery~\cite{DeFinery}} is the first tool to incorporate formal verification for patch validation. Similarly,  \textbf{SmartFix~\cite{SmartFix}} combines mutation with formal verification \cite{verismart} and stands out by integrating a statistical model to prioritize likely correct patches.

A second family of approaches involves applying templates that are tailored for different vulnerabilities.
\textbf{sGuard~\cite{sGuard}} was the first contribution in this direction.
\textbf{Aroc~\cite{Aroc}} creates on-chain fixes with templates, we exclude it from the rest of this paper, due to its reliance on a modified EVM for repair.
\textbf{HCC~\cite{HCC}}, the Hardening Contract Compiler, applies templates to add security checks at compilation time, returning a source code patch. 
\textbf{ContractFix~\cite{ContractFix}} uses a multi-level template approach that operates at three levels of granularity: statement level, method level, and contract level. 
\textbf{TIPS~\cite{TIPS}} proposes a multi-template strategy where a vulnerability type can be fixed by multiple templates.
\textbf{sGuard+~\cite{sGuardPlus}} takes the problem of accurately localizing the vulnerability by connecting bytecode, source code, and AST representation of the vulnerability in a contract. 
\textbf{vFix~\cite{fang2024vfix}} enhances the template applicability through pointer analysis and intra-procedural data-flow analysis. 
Finally, \textbf{ReenRepair~\cite{ReenRepair}} focuses on reentrancy vulnerability and attempts to produce gas-optimized patches by reordering source code statements that correlate a READ and WRITE instruction in the bytecode. 

Two APR tools for smart contracts use supervised learning: 
\textbf{SmartRep~\cite{SmartRep}}  employs a double encoder to abstract the vulnerable method from the AST, and train a system from one-line fixes of GitHub commits.
\textbf{RLRep~\cite{RLRep}} applies a reinforcement learning approach with policy gradient, using a reward function that evaluates fixes based on compilation success, vulnerability detection, code entropy \cite{entropy}, and code similarity.

The latest trend in APR for smart contracts involves using Large Language Models (LLM). 
\textbf{GPT\&BARD~\cite{gpt&bard}} tests \texttt{gpt-3.5} under three scenarios: vulnerable code with repair examples, solely vulnerable code, and vulnerable code with static analysis output.  
\textbf{SolGPT~\cite{SolGPT}} uses \texttt{gpt-3.5-turbo}, combining vulnerable code with static analysis results from Slither\cite{slither}.  \textbf{ACFix~\cite{zhang2024acfix} } addresses access control vulnerabilities by generating a role-pair taxonomy and using \texttt{gpt-4} agents for generation and validation. 
\textbf{ContractTinker~\cite{contracttinker}} applies an agentic approach using \texttt{gpt-4} to define a vulnerability mitigation plan and repair action.

Finally, there is an original line on binary APR tools, which repair EVM bytecode directly. 
\textbf{SmartShield~\cite{smartShield}} extracts context from a control-flow-graph (CFG) and operates at the bytecode level to address missing checks on insecure operations and state changes after external calls. 
\textbf{EVMPatch~\cite{EVMPatch}} apply repair on EVM bytecode with a trampoline technique, which is a technique to add new EVM instructions into empty code areas.
Lastly, \textbf{Elysium~\cite{Elysium}} applies context-aware patches using semantic-based techniques to gather relevant information like integer type and free storage space.
All those binary tools are template-based approaches.

\subsection{Detection \& Repair Level}\label{subsec:level}

When studying program repair tools for smart contracts, the first classification to make is the level at which the tool operates. We identify two categories: source code level and bytecode level tools. Some tools can mix both in detection and/or repair. 

The source code level refers to textual files in high-level programming languages (Solidity in this case, but could apply to other smart contract languages). 
The bytecode level refers to low-level instructions generated after compiling the smart contract source code. In this paper, it is specifically the EVM bytecode that is being considered.

\textbf{Detection.} When applying vulnerability detection on source code, the goal is to identify vulnerable lines of code. A potential downside of source code detection is the rate of false positives as the detectors fail to identify genuine problems in the code \cite{ChaliasosDeFiTools}.

\revision{Vulnerability detectors at the bytecode level identify a particular instruction location in the EVM binary code.} The benefit of detecting at the bytecode level is that it comes after compilation and optimization, so it reduces the false positive rate happening before compilation. A downside of detecting at this level is the lack of understandability of the binary code as it is not human-readable, and the difficulty mapping back to source code.

\textbf{Repair.} Repairing at the source code level is advantageous because it enables a developer to assess the quality of a patch with manual inspection. A downside is the difficulty of abstracting vulnerabilities in certain repair patterns as code can be written in thousands of different ways.

A common challenge of repairing at the bytecode level is the lack of high-level contextual information, including names and identifiers,  lost during compilation.

Out of 19 selected tools, 16 repair at the source code level, and 3 tools apply repair at the bytecode level.
We see that tools working on repair at the source code level are more popular. We believe this is due to the necessity of proposing patches that a human can assess and validate.

\subsection{Vulnerability Detection} \label{subsec:vulnerabilitydetection}
The first phase of an APR tool is localizing the vulnerability it aims to repair. Based on our literature review, we categorize vulnerability detection into 3 groups:

\textbf{User-provided (U-P):} The tool asks the user to categorize the contract vulnerabilities' according to a predefined format. For example, listing~\ref{lst:UP-TIPS} presents the format to follow for User-Provided detection on TIPS~\cite{TIPS}.

\textbf{External (EXT):} The tool relies on an external detector. 
For example, ContractFix~\cite{ContractFix} reuses Slither~\cite{slither}, Securify~\cite{securify} and Smartcheck~\cite{smartcheck}.

\textbf{Internal (INT):} The APR tool has its own mechanism for vulnerability detection. In this case, the tool contribution is both on the repair side and the detection side. An example of this category is sGuard~\cite{sGuard} which proposes a new detection model and repair strategy.

\begin{center}
    \footnotesize
    \begin{lstlisting}[language=json , caption={ User-Provided  vulnerability detection information for TIPS \cite{TIPS} } , float, label={lst:UP-TIPS}, xleftmargin=0.5cm]
"name": "FibonacciBalance.sol",
"defect": [ {
    "lines": [ 31, 38  ],
    "category": "access_control" }]
\end{lstlisting}
\end{center}

\revision{We found that 5 tools use an internal detection mechanism, 3 rely on user-provided detection, and 12 use external detection.} Also, source code tools that use external sources tend to rely on Slither~\cite{slither} (7/10), and byte-code tools on Osiris~\cite{osiris} (3/3).

\subsection{Repair Strategy } \label{subsec:repairstrategy} 
The main component of program repair tools is the one responsible for patch generation. 
The literature on repair for smart contracts relies on the following techniques.

\textbf{Code Mutation:} (\cite{SCRepair,DeFinery,SmartFix}) This involves performing changes on the code according to specific transformation operators, in order to find a solution to a vulnerability. The scope for mutations can widely vary and therefore there is little overlap between the mutations considered in each paper.

\textbf{Template:} (\cite{sGuard,Aroc,HCC,ContractFix,TIPS,smartShield,EVMPatch,Elysium,ReenRepair,sGuardPlus,fang2024vfix}) This mechanism consists of creating templates designed for specific vulnerabilities. Tools can consider one template per vulnerability or many per family type. The second has been proven to improve repair capabilities~\cite{TIPS}.

\textbf{Supervised Machine Learning:} (\cite{RLRep,SmartRep}) \revision{Predicts vulnerability patches based on previous fixes.} A common challenge is the lack of such training data for smart contracts (labeled fixes) \cite{RLRep}.

\textbf{Generative Machine Learning (LLM):}  (\cite{gpt&bard,SolGPT,zhang2024acfix,contracttinker}) LLMs are trained on vast amounts of unsupervised data. It has been shown that they can be applied to smart contract coding tasks, including program repair. This approach involves feeding an LLM with source code and a vulnerability to get a patch.
A challenge of this approach is the absence of logical explanation about the patch, with no logical reasoning about the process to come to the fix~\cite{zhang2024acfix}
        
To sum up, the most popular technique in the literature is template-based, yet learning-based approaches have emerged in more recent studies.

\subsection{Patch Validation}\label{subsec:patchvalidation}
After generating patches, the next phase in APR is to validate them. Validation may involve ensuring functional correctness and/or vulnerability mitigation. This study focuses on vulnerability mitigation. 
In our survey of APR for smart contracts, we identified the following approaches for ensuring vulnerability mitigation:

\textbf{Static Analysis (SA):} The dominant technique in APR for smart contracts is running a static analyzer tool on the generated patches to statically check for vulnerability mitigation. 
SolGPT~\cite{SolGPT}, for example, validates patches using Slither~\cite{slither} to verify that the vulnerability has disappeared. The drawback of this technique is its reliance on the static analyzer's accuracy, as any false positive or false negative will affect the validation step of a tool.

\textbf{Formal Verification (FV):} Formal verification offers a deductive approach to analyzing the vulnerabilities in a program. 
It consists of mathematically proving the presence or absence of a vulnerability in a contract. 
In our study, DeFinery~\cite{DeFinery} and SmartFix~\cite{SmartFix} are the only tools that utilize this technique in the context of smart contract security.

\textbf{Manual Evaluation (ME):} Using their expertise, some authors manually determine if the vulnerability is still present in the patched code. 
An example of this validation technique is sGuard+~\cite{sGuardPlus}. 
The drawbacks of manual evaluation are the lack of scalability and reproducibility.

Automated repair of vulnerabilities should work only with true positive vulnerabilities in a contract. 
True positive vulnerabilities mean that exists an exploit that can be executed to the contract~\cite{vulnotexploit}. 
To validate patches for vulnerability mitigation, one should use each corresponding exploit.

In RQ3~\ref{subsec:rq3-methodology} we propose an innovative validation technique by generating exploits for the vulnerable contracts in our study.

\subsection{Recapitulation }
To sum up, past contributions on APR for smart contracts exhibit considerable variations in their detection, repair, and validation methods. The \totaltools tools analyzed in our study rely on different detectors, target distinct vulnerability categories, and are evaluated on diverse, often incompatible benchmarks.
It is impossible for the research and practitioner communities to truly understand where the state of the art is, and what the main strengths and weaknesses of each tool are.
Our work addresses this gap by empirically testing APR tools on a common dataset and evaluating their ability to mitigate actual exploits in a sound manner.

\section{Experimental Methodology}
\label{sec:experimental_methodology}

\subsection{Research Questions}
We devise a novel experimental methodology to study APR for smart contracts.
\revision{Our study is articulated around five research questions.}

\begin{itemize}
    \item RQ1 (scientific reproducibility): \rqone
    \item RQ2 (systematic comparison): \rqtwo
    \item RQ3 (exploit mitigation): \rqthree
    \item RQ4 (efficacy per vulnerability type): \rqfour
    \item \revision{RQ5 (manual analysis): \rqfive}
\end{itemize}

\subsection{RQ1: Scientific Reproducibility}
\label{subsec:rq1-methodology}

After consolidating the list of APR tools to assess (Table~\ref{tab:ToolDescription}), we proceed to systematically test all tools for basic executability and reproducibility.
This experiment has the following steps:
\begin{enumerate}
    \item \textbf{Availability}: For all papers, we search for the repository link in the article.
    When having an unsuccessful search, we reach out to the authors asking for the code.

    \item \textbf{Installability}: We try to set up the environment, install the dependencies, and compile, if applicable, the tool's source code, per the provided documentation.
    When struggling with this step, we reach out to the authors asking for further instructions. 

    \item \textbf{Executability}: \revision{Finally, if the previous steps succeed, we test the tools by executing them using both the example inputs provided in their repositories and custom test contracts.}
\end{enumerate}

In cases where a step failed, our initial approach was to manually debug and resolve the errors, keeping the modifications as minimal as possible.
Before deciding to drop a tool from further experiments, our last attempt is always to directly communicate with the authors.
 
\revision{This experiment provides an overview of reproducibility in smart contract APR, showing that results are reproducible within the tools' claimed operating conditions. 
It also identifies a set of executable APR tools that can be used for deeper investigation in subsequent research questions.}

\subsection{RQ2: Basic Repair Effectiveness} \label{subsec:rq2-methodology}

In this experiment, we aim to run all reproducible APR tools (see Table~\ref{tab:RQ1results}) on a common benchmark. To our knowledge, this has never been done before.
As a benchmark, we use the well-known \dataset dataset~\cite{SmartBugs}, which has been widely employed in related research. The dataset comprises 143 vulnerable smart contracts, each annotated with one of ten categories derived from the DASP taxonomy:
\href{https://dasp.co/}{reentrancy}, \href{https://dasp.co/#item-2}{access control}, \href{https://dasp.co/#item-3}{arithmetic}, \href{https://dasp.co/#item-4}{unchecked low-level calls}, \href{https://dasp.co/#item-5}{denial of services}, \href{https://dasp.co/#item-6}{bad randomness}, \href{https://dasp.co/#item-7}{front running}, \href{https://dasp.co/#item-8}{time manipulation}, \href{https://dasp.co/#item-9}{short addresses}, and \href{https://dasp.co/#item-10}{miscellaneous}. \dataset is well appropriate for our study because its widespread use in previous research provides a foundation for comparison. Importantly, all the tools we evaluate are fully compatible with this dataset. This compatibility ensures a consistent evaluation framework and facilitates a fair comparison across tools, making \dataset well-suited for our experiment. 

\revision{Listing~\ref{lst:dataset-example-reentrancy} shows an example of a contract from the dataset with the reentrancy vulnerability in its \texttt{withdrawBalance()} function. On line 6, the contract sends funds via \texttt{call.value()} before zeroing the user's balance on line 9. This critical sequence allows an attacker to exploit the code by recursively calling the \texttt{withdrawBalance()} function through a fallback function, repeatedly withdrawing funds before the balance is updated, effectively draining more funds than their initial balance.}

\input{solidity/contract_reentrancy}

On top of the existing benchmark infrastructure, we add a novel evaluation framework to assess program repair capability and patch validity. 
This patch assessment does not exist in \dataset. 
The patch assessment framework comprises five steps. Each step is described in detail below.

\begin{enumerate}
    \item \textbf{Detection Accuracy}: We assess each tool's ability to detect the original vulnerability. \dataset uses the DASP taxonomy to classify vulnerabilities.
    \revision{For each vulnerability category detected by the tools, we manually match it to the appropriate DASP class by reviewing the tool's description.}
    Then, we compare the ground truth DASP class against the detected one on each contract.
       
    \item \textbf{Patch Generation:} Each tool is configured according to its original instructions to generate patches. For each contract in the dataset, we generate and collect all patches produced by the tools. 

    \item \textbf{Patch Compilation:} We try to compile all the patches from the previous step using the appropriate compiler version. This step assesses the syntactic and typing correctness of the produced patches. Note that this step does not apply to binary APR tools.

    \item \textbf{Differential Analysis:} We note that some tools sometimes do not change anything in the original contract. To rule out ineffective repairs, we compare each generated patch against the original contract file. This involves examining the diff files to ensure the patches effectively modify the original code. 
    
    \item \textbf{Consistency via Tool Detector:} 
    Finally, we check whether the patch fixes the original vulnerability according to the tool's own detector.
    We compare each tool's vulnerability report of the original smart contract and its respective patch and check that the vulnerability is no longer detected. \revision{This means that the tool's detection and repair algorithms are consistent with each other.}
\end{enumerate}

\subsection{RQ3: Exploit Mitigation Rate}\label{subsec:rq3-methodology}
Next, we evaluate the effectiveness of the considered tools in truly mitigating smart contract exploits.

As previously mentioned in \autoref{subsec:patchvalidation} Patch Validation, automated program repair for smart contracts aims at preventing exploits while maintaining functionality.
Therefore, evaluating a patched contract's ability to mitigate an exploit is a strong criterion for repair effectiveness, stronger than, for example, the absence of statically detectable issues.
To our knowledge, this concept of performing actual exploit mitigation has never been done at this scale.

There is a fundamental challenge to doing this: the absence of executable exploits.
Existing vulnerability datasets, including \dataset, have vulnerabilities but no exploits.
In this experiment, we first manually write successful exploits for our dataset and then run the exploits on the patched contracts to check whether the patch mitigates them.

\begin{figure*}
    \centering
    \includegraphics[width=1\linewidth]{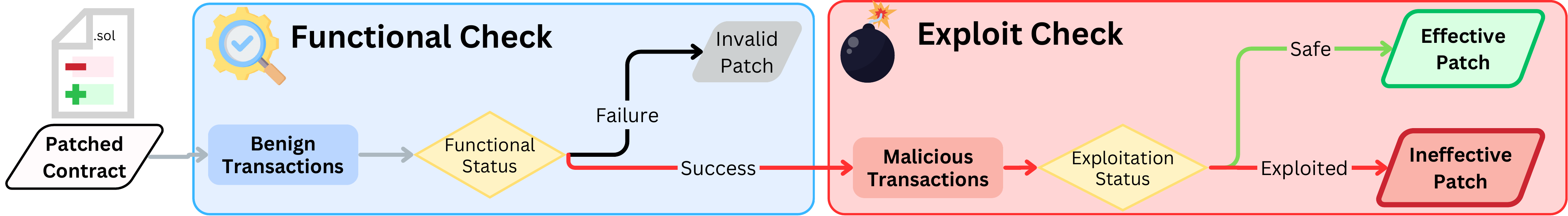}
    \caption{RQ3: Our novel methodology for evaluation via actual, executable exploits, a core contribution to the field of smart contract repair.}
    \label{fig:rq3_methodology}
\end{figure*}

\textbf{Exploit Writing.}
We manually write executable exploits for \dataset dataset.
Each exploit targets a specific vulnerability and is verified by violating a contract's core property: either safety properties (e.g., unauthorized token transfers) or liveness properties (e.g., denial of service).
The manual creation of each exploit is time-blocked for 1 hour.
We identify the attack vector by understanding the contract's logic, interface, and the labeled security weakness.
We construct a precise transaction sequence that reliably exploits the vulnerability.
When necessary, our exploits incorporate auxiliary smart contracts that are called attacker contracts, enabling the attack.


We propose an original methodology to assess exploit mitigation,  illustrated in Figure~\ref{fig:rq3_methodology}. This methodology has two main steps: Functional Check and Exploit Check.

\begin{enumerate}
    \item \textbf{Functional Check:}
We note some patches repair vulnerabilities at the cost of breaking important contract functionality.
For each exploit, we create multiple benign transactions that exercise the function without triggering the vulnerability.
\revision{These are manually created by analyzing the contract's source code and associated metadata (such as historical transactions, if any). The benign transactions are designed to safely exercise the vulnerable code paths according to the contract’s intended behavior, they do not test the full contract functionality. The correctness of these transactions is cross-checked between authors.}
These transactions must always succeed on a patched contract, confirming that the patch preserves basic functionality.
If any benign transaction fails, the patch is considered invalid.
To sum up, functional checks verify that the patched contract maintains its intended behavior for benign use cases.

This check is essential because the exploit harness only verifies attack success or mitigation, but does not detect if the patch breaks the contract.
From a methodological standpoint, without functional checks, such broken patches can be misinterpreted as effective. Our work is the first to address this key methodological aspect in program repair for smart contracts.

\item \textbf{Exploit Check:}
We use the Hardhat Framework~\cite{hardhat} to ensure consistent execution and reproducibility of these exploits;
each exploit is executed as a Hardhat test.
The test orchestrates a structured sequence of transactions on a local blockchain.
These transactions deploy the smart contracts and trigger their respective vulnerabilities.
Finally, the test performs a verification step using pass/fail condition checks to confirm whether the exploit successfully compromised the contract.
\revision{In other words, to check safety properties, assertions are added to detect invalid program states and violations of invariants. For liveness, the benign transactions ensure no deadlocks or interruptions of normal operations.
}
\end{enumerate}

To answer RQ3, we focus on the following metrics: functional check failure rate and exploit mitigation rate.
The failed functional check rate indicates the percentage of patches that break the contract's functionality.
This helps identify patches that render the contract unusable for its designed purpose.
The exploit mitigation rate measures the proportion of patches that successfully achieve two goals: they prevent the original vulnerability from being exploited while maintaining the contract's core functionality.
This metric provides direct insight into how effectively each tool can generate patches that improve security without introducing functional regressions.

\revision{We note that none of the exploit generation tools, such as \texttt{teEther}~\cite{teEther}, are capable of generating those exploits because they focus on very specific attacks that are not considered in the dataset used in this study.}

\subsection{RQ4: Efficacy per Vulnerability Type} \label{subsec:rq4-methodology}
Following, we aim to analyze the efficacy of vulnerability repair from the perspective of vulnerability types. 
\revision{
To address RQ4, we aggregate the results of RQ3 per vulnerability type, using the metadata provided in our dataset.
This analysis examines exploit mitigation from two perspectives: the types of mitigated vulnerabilities and the performance of each tool with respect to the vulnerabilities it is explicitly designed to address.}

 \revision{This research question helps identify the vulnerability types effectively addressed by current tools and those that still require significant attention from future research.}
\begin{rblock}
    
\subsection{RQ5: Manual Analysis of Patches}

To deepen our understanding of patch quality, we conduct a manual evaluation of source code patches that mitigate an exploit in our dataset using two sampling strategies:

\textbf{1) Tool-based Sampling:} From the patches that successfully mitigate exploits per the automated check, we randomly select 10 samples produced by each source code level tool.

\textbf{2) Vulnerability type based Sampling:} To assess repair efficacy across vulnerability classes, we randomly selected 10 exploit-mitigating patches, if available,  for each vulnerability type considered.

Each selected patch is manually examined to assess two key criteria: (1) whether it disrupts the expected normal behavior of the contract, and (2) whether it effectively mitigates the underlying vulnerability. 
The manual review is independently conducted by two authors, with disagreements resolved through discussion with a third author.

Additional factors considered during the evaluation include the correctness of the associated exploits and the potential introduction of new vulnerabilities.
\end{rblock}

\section{Experimental Results}
\label{sec:experimental_results}
\subsection{RQ1: Scientific Reproducibility}

Following the methodology outlined in Section~\ref{subsec:rq1-methodology}, we present the findings from our reproducibility experiment. Out of an initial set of \totaltools tools, \toolsavailable tools have publicly available source code, while the source code for the remaining 4 tools was inaccessible. The reasons were copyright restrictions (2 cases) or the absence of a response from the authors (2 cases). Table~\ref{tab:RQ1results} summarizes our efforts in installing and executing these \toolsavailable tools. We successfully installed \totalinstall tools using their source code. We encountered issues such as missing or deprecated libraries with the remaining ones.
\revision{This is the case for SCRepair and ContractFix where required Python modules and NPM packages, respectively, were not available in the source code.
DeFinery was not reproducible due to failure in installing its dependencies.
}

\revision{Among the \totalinstall installed tools, \totalexecutable were successfully executed according to their provided usage examples and additional smart contracts, while 2 failed during execution due to runtime errors or crashes, and 2 could not be executed due to insufficient documentation and unclear input requirements.}
Notably all tools employing machine learning techniques failed the training execution.
\revision{SmartRep execution fails during the training process, where it encounters an unexpected exception.
RLRep fails in loading its necessary modules and the training process fails to update the model.
vFix lacks sufficient documentation or examples on how to use detection reports to run the tool, making it infeasible to execute.
ContractTinker successfully runs on the provided examples, yet we could not generalize its execution to other smart contracts due to missing documentation on required input formats, particularly regarding the expected structure of vulnerability reports.}

Overall, we were able to execute \totalreproducible /\totaltools research tools from the literature. For them, we only had to make minor modifications to the code to address syntax errors or dependency issues.

Our results underscore challenges in research transparency (one-fourth of the tools presented in peer-reviewed articles lack publicly available code).
Even when the code is available, there are serious reproducibility issues, as less than half were reproducible.

To improve this situation, we make the following recommendations.
1) Open-sourcing code with clear usage instructions, ideally on a platform like GitHub that supports version control and collaboration.
We believe that GitHub facilitates error resolution compared to Anonymous GitHub or Zenodo.
2) Adhering to good coding practices, such as including a build file with dependencies and their versions.
3) For machine learning tools, provide the fully trained models alongside the training source code, as the training process is resource-intensive and prone to failure.

\input{tables/RQ1/results}

\begin{answerbox}{Answer to RQ1} \textbf{\rqone}\\
From an initial dataset of \totaltools APR tools, \toolsavailable have publicly available source code, \totalinstall can be successfully installed, and \totalreproducible can be successfully executed end-to-end with only minor changes to the provided code.
\end{answerbox}

\subsection{RQ2: Basic Repair Effectiveness} \label{ssec:rq2-results}

Table~\ref{tab:over_comparison_rq2} shows the result of our large scale repair experiment on dataset \dataset per our evaluation methodology described in \autoref{subsec:rq2-methodology}. 
The columns correspond to the different steps of the methodology. The columns are Detected, Generated, Compilable, Different, and Consistent.
In this table, column C stands for the number of vulnerable contracts with at least one patch that fulfills the criterion of the column, and P stands for the number of patches that satisfy the same criterion.

\input{tables/RQ2/results}

\textbf{Detected C.} Following our methodology we first check whether each tool detects the vulnerability according to the ground truth from the dataset. 
The accuracy varies considerably. SolGPT achieves the highest accuracy, by detecting 97/\totalcontracts vulnerable contracts. In contrast, sGuard only identifies 35/\totalcontracts contracts as vulnerable, the lowest accuracy among all tools. 
 
On Table~\ref{tab:over_comparison_rq2} we found for 119 contracts there is at least one tool correctly detecting the label vulnerability. It confirms that SolGPT has the best strategy for detecting the vulnerability.
Also, we note that there is an overlap among detection. This suggests that an ensemble method for vulnerability detection may be valuable for wider coverage.
Since no tool detects 24 contracts as vulnerable, the dataset \dataset is not yet exhausted for research.

\textbf{Patch Generation.}
The second metric of the study focuses on patch generation. The dataset includes \totalcontracts contracts. TIPS achieves 140 patched contracts, followed closely by SolGPT with 139. Notably, SolGPT and TIPS are the only tools able to generate multiple patches per contract. This leads to SolGPT producing 552 patches and TIPS generating 242 patches over the whole dataset, more than the total number of contracts. 
One contract (parity\_wallet\_bug\_1.sol) did not obtain a single patch from any tool due to a solidity version required by the contract but incompatible with the experimental pipeline.

\textbf{Compilable.}
For 142 out of \totalcontracts contracts, at least one patch compiles. sGuard+ and SmartFix achieve a 100\% compilation success rate.  The remaining three tools produce at least one non-compilable patch.
The non-compilable patches in sGuard are due to a syntax error caused by an incorrect variable in the code. For SolGPT, some patches fail due to syntax errors stemming from non-code elements accidentally included in the LLM response, which is a known problem for LLM-based code generation. In TIPS, compilation fails for only two smart contracts, both due to syntax errors related to improper handling of string literals.
Overall, the compilation results are good, all tools have a high ratio of compilable patches, showing reasonable engineering quality.

\textbf{Different.}
Our fourth step in this experiment is to run a differential analysis. With this metric, we ensure that the produced patches actually modify the original code. sGuard+, SmartFix, SolGPT, and TIPS always produce patches different from the original contract, as expected. In contrast, sGuard achieved only \revision{57\%} unique patches, with 46 contracts remaining identical to the original code. Similarly, Elysium produced \revision{79\%} different patches, leaving 26 unchanged contracts, while SmartShield reached 100\% uniqueness. Our results indicate the relevance of this metric, which is often overlooked in related research. 

\textbf{Consistent.} 
Finally, we evaluate the consistency between the tool's detector and the patch generator. This means if the tool considers the vulnerability, initially detected in the contract, as disappeared for the generated patches. A patch is considered consistent if all instances of the detected vulnerability are undetected after patching. A contract is considered consistent if there is at least one patch that removes all instances of the detected vulnerability. 
Among the tools, SolGPT is consistent for 88/97 contracts, the highest absolute number across tools. TIPS followed closely with 81/82 consistently fixed contracts, while SmartFix achieved 48/51.
sGuard+ gathered 70/70 consistent contracts achieving 100\% consistency on the detected contracts. 
sGuard only achieves consistency for 2/35 contracts.
\revision{
This discrepancy stems from differences in the tools’ detectors.
Specifically, sGuard’s detector does not account for addition of protective code and continues to flag vulnerabilities based on low-level patterns, such as unsafe opcode sequences.
Although sGuard and sGuard+ share similar template-based repair approaches, their detectors differ substantially, leading to distinct outcomes.
These findings suggest that adopting more sophisticated and context-aware detectors can significantly improve consistency.}


\begin{answerbox}{Answer to RQ2} \textbf{\rqtwo}\\
Our evaluation shows that 142 contracts have multiple patches from diverse APR tools.
Yet, we observe inconsistencies in detection accuracy and patches which severely undermine soundness. These findings highlight the urgent need for better methodologies, incl. robust patch validation, and reliable end-to-end repair pipelines.
\end{answerbox}

\subsection{RQ3: Exploit Mitigation Rate}\label{subsec:rq3-results}
We successfully manually create 91 exploits for the \dataset dataset. Each exploit targets one vulnerable contract in the dataset. For every exploit, we also write a robust, handwritten functional check to evaluate the functionality preservation of the patched contract.
Table~\ref{tab:RQ3results} summarizes the results of our exploit experiments conducted on 91 contracts with known vulnerabilities.

\textbf{Failing Functional Check Rate.}
Table~\ref{tab:RQ3results} presents the functional failure rates for each tool. The results reveal that certain patches disrupt the core functionality of the contracts.

Our findings show that 6 out of 7 tools fail to preserve functionality in some contracts, emphasizing the critical importance of functional checks when assessing smart contract patches.
\revision{Our results reveal a clear dichotomy between bytecode-level and source-code-level repair tools.
Bytecode tools (Elysium, SmartShield) exhibit significantly higher functional failure rates (26/91 and 21/91, respectively) compared to source-code tools (average 3/91). This suggests that operating at the bytecode level makes it substantially harder to preserve the original contract logic and functionality.
While bytecode repair offers the advantage of working without source code, our findings indicate this comes at the cost of frequently breaking core contract functionality.}

In contrast, template-based tools such as sGuard, sGuard+, and TIPS exhibit high functionality preservation, with low failure rates. This suggests that their template-based repair strategies are more effective in maintaining contract functionality. However, these templates are not universally applicable, and some patches still lead to functional breakdowns.

\input{solidity/tips_patch_diff}

\input{solidity/smartfix_patch_diff}

For instance, Listings~\ref{lst:invalid_patch} and~\ref{lst:effective_patch} both address reentrancy exploits but yield differing outcomes. Listing~\ref{lst:invalid_patch} illustrates a TIPS patch that mitigates reentrancy by setting the user's balance to zero before transferring it. While this approach prevents the exploit, it also renders the \texttt{withdrawBalance} function unusable, as it always transfers zero. Conversely, Listing~\ref{lst:effective_patch} showcases a SmartFix patch for the same vulnerability. This patch stores the user's initial balance, updates it, and then transfers the stored value, successfully preventing reentrancy while preserving functionality.
This example highlights the importance of functional checks in detecting patches that disrupt contract functionality, which the exploit harness alone cannot identify.
Without functional checks, such broken patches could be misclassified as effective.
\input{tables/RQ3/mitigated_exploits_by_tool}

\textbf{Exploit Mitigation Rate.}
Our exploits are attempted on every patched contract, with the corresponding results shown in Table~\ref{tab:RQ3results}.
The effectiveness in mitigating exploits varies significantly.
SolGPT achieves the highest mitigation rate with 67 patches (\revision{74\%} of \exploitdatasetname) mitigating the exploit, while SmartShield has the lowest with \revision{26} exploits mitigated, representing \revision{29\%} of the exploit dataset.
There is a clear disparity in the tools' ability to effectively address smart contract vulnerabilities.
We note that no single tool successfully mitigates all exploits, there are still 11 contracts with an exploit that no tool succeeds in mitigating.

The third column of Table~\ref{tab:RQ3results}, ``Unique'' shows the exploits that only one tool can handle uniquely.
SolGPT and TIPS mitigate the highest number of unique exploits (4), followed by SmartFix and Elysium, each with 1 exploit not covered by others.
This analysis highlights the necessity for diverse mitigation strategies.
We believe that industrial tools for smart contract repair will follow an ensemble based repair approach to maximize effectiveness.

\revision{
Our results demonstrate that no single repair tool is universally optimal. While SolGPT achieves the highest mitigation rate (74\%), template-based tools like TIPS best preserve functionality (only 1 failed functional check), and bytecode-level tools (e.g., SmartShield) offer the advantage of source-free repair but at the cost of higher failure rates (28\%). This suggests that practitioners should not rely on a single tool but instead adopt ensemble strategies.
For instance, they can prioritize template-based patches if they are available, and only use LLM-generated fixes for cases not covered by templates.
Future work should 1) explore hybrid techniques to address the 11 still-unmitigated exploits and 2) improve bytecode-level repair, as functional correctness remains a critical challenge.
}

\begin{answerbox}{Answer to RQ3} \textbf{\rqthree}\\
This is the first ever experiment to demonstrate that APR tools for smart contracts actually mitigate smart contract exploits.
SolGPT stands out with the highest mitigation rate, stopping \revision{74\%} of the exploits.
Next comes SmartFix, a mutation-based tool, which mitigates \revision{53\%} of the exploits.
Template-based tools such as TIPS (49\%), sGuard+ (48\%), and sGuard (\revision{33\%}) show moderate mitigation success but are notably better at preserving smart contract functionality.
Our publicly available dataset of functional checks and exploits will help future researchers in conducting a strong, sound evaluation of APR for smart contracts.
\end{answerbox}

\subsection{RQ4: Efficacy per Vulnerability Type } \label{ssec:rq4-results}

\revision{The results of our analysis of mitigation effectiveness by vulnerability type are presented in Table~\ref{tab:rq4_results}. For each vulnerability type in the dataset, the table reports the number of exploits in our sb-heists dataset, the number of mitigations per tool, as well as the aggregated mitigation counts and overall relative effectiveness. On each tool's column, ``n/a" stands for not applicable when the tool's paper or documentation does not declare that it addresses that particular vulnerability type.}

\textbf{Unchecked Low-Level Calls.}
This is the most prevalent vulnerability type in the dataset, with 52 contracts and 20 exploits. \revision{Of the 7 reproducible tools in our experiment, 5 explicitly target this vulnerability.} Our results show that 19/20 contracts have a patch that mitigates the unchecked low-level call vulnerability. Overall, this means that 95\% of such vulnerabilities are properly mitigated, meaning that this problem can be considered solved.

\textbf{Reentrancy.}
In \dataset, 31 contracts belong to the reentrancy family of vulnerability, 26 of them have a corresponding executable exploit in \exploitdatasetname. \revision{All 7 tools evaluated in this study are designed to address this vulnerability.} 
Our result indicates that for all reentrant contracts, there is at least one patch that mitigates the exploit, which is arguably a strong result. This is an encouraging result for smart contract developers, who can use already detection and repair tools to handle reentrancy.

\textbf{Access Control}
The access control vulnerability appears in 18 contracts in \dataset, with 16 associated exploits. \revision{In this study, 6 out of the 7 tools are designed to target this vulnerability type. Our results show that for 13 out of 16 exploits, there is at least one patch that successfully mitigates the issue, corresponding to a repair effectiveness of 81\%}. While this indicates meaningful progress, it also highlights the need for further research to fully address access control vulnerabilities.

\textbf{Arithmetic}
This vulnerability category includes 15 contracts in the dataset, with 13 corresponding exploits. \revision{In this study, 6 out of the 7 reproducible tools are designed to address this category.} Arithmetic vulnerability patching has a perfect 100\% effectiveness. In addition, we note that this has also been fully handled at the compiler level, in release 0.8 of the Solidity compiler.
 
\input{tables/RQ4/results}

For the rest of the vulnerabilities in the dataset, we have a reduced number of contracts and corresponding exploits. For \textbf{Bad Randomness}, there are 4 exploits and a 100\% effectiveness for exploit mitigation, \textbf{Time Manipulation} got \revision{67\%} mitigation effectiveness over the 3 exploits, and  \textbf{Front Running} one single mitigated exploit (33\%).
\revision{On the other hand, for \textbf{Denial of Service} no tool could produce patches for this vulnerability.} This calls for further research on these vulnerability types.

 \revision{As shown in Table~\ref{tab:rq4_results}, APR tools work well on Unchecked Low Level Calls, Reentrancy, Access Control, and Arithmetic vulnerabilities, with all those vulnerability types having an effectiveness higher than 80\%.
 None of the tools explicitly target Denial of Service and Front Running.}
 
 \revision{However, this does not necessarily imply zero patch generation, as demonstrated by Elysium and SmartShield, which produced patches for Front Running. Since these tools provide bytecode patches, further analysis is out of scope for this study.}

Overall, \revision{88\%} of exploits are mitigated at least once, showing that research has made significant progress towards automatic mitigation of smart contract exploits. Yet, we bear in mind that the contracts in \dataset do not reflect the full complexity of real-world contracts.

\begin{answerbox}{Answer to RQ4} \textbf{\rqfour}\\
Our study of exploit mitigation rate per vulnerability type demonstrates high effectiveness for most of them. Reentrancy, Arithmetic, and Bad Randomness exploits can all be mitigated. Unchecked Low-Level Calls and Access Control exploits are also mitigated with high effectiveness rates of 81\% and 95\%, respectively. These positive results show that the research community has made good progress, but also hint that \dataset may start to be an exhausted benchmark. Our results clearly motivate the construction of new benchmarks with 1) more vulnerability types (e.g. price manipulation vulnerabilities and 2) more real-world smart contracts.
\end{answerbox}

\begin{rblock}
    
\subsection{RQ5: Manual Analysis of Patches } \label{ssec:rq5-results}
To assess the quality of the generated patches, we manually analyze a total of 111 patches. These include ten randomly selected patches per tool that produce source code fixes, as well as ten (when available) per vulnerability type.

All 111 patches successfully block the corresponding exploit, providing strong evidence for the construct validity of our automated validation framework.
However, three patches break legitimate functionality, revealing the challenge of designing comprehensive functional checks.

Two of the incorrect patches, applied to contracts with unchecked low-level call vulnerabilities, alter the original contract behavior under specific conditions.
These contracts rely on external contracts not included in the \dataset dataset.
Depending on the external source code, the patch may or may not introduce behavioral changes.
The third incorrect patch targets an access control vulnerability in a proxy contract.
In this case, the tool replaces a critical delegatecall with a simple call, which mitigates the vulnerability but alters the intended behavior, effectively bypassing the functional check.
All three cases go undetected by the automated framework due to incomplete sanity checks.

An interesting case is that of ``Bad randomness'', where all reviewed patches mitigate the exploits by changing the hash function used for randomness. However, the new computation still follows the same insecure pattern~\cite{bad_randomness}, meaning that while the concrete exploit is blocked and the functional checks pass, the general problem remains. 
This highlights a key limitation of exploit-driven validation: although patches may successfully block concrete exploits, they may overlook the broader vulnerability class, particularly in cases where the exploitability depends on subtle execution characteristics.

We note that repair tools, especially learning-based ones, may introduce new vulnerabilities while fixing the target one.
We explicitly check for such cases in our manual analysis of 111 patches.
None of them introduces a new vulnerability. 

\begin{answerbox}{Answer to RQ5} \textbf{\rqfive}\\
\revision{
Out of 111 manually evaluated patches, 108 (97\%) are found to be logically correct by three expert reviewers, demonstrating the reliability of our automated validation framework. 
Only three patches are deemed incorrect, with each failure being explained by corner-case limitations in our functional check mechanism; calling for future work on exhaustive functional checks.}
\end{answerbox}
\end{rblock}

\section{\revision{Discussion}}
\label{sec:discussion}
\begin{rblock}
Our empirical evaluation of smart contract APR tools reveals critical insights about their capabilities, limitations, and directions for future research.
We organize our discussion around three key themes that emerge from our empirical results.

\subsection{Impact of Detection Accuracy on Repair Performance}
\begin{figure}[ht]
    \centering
    \includegraphics[width=\linewidth]{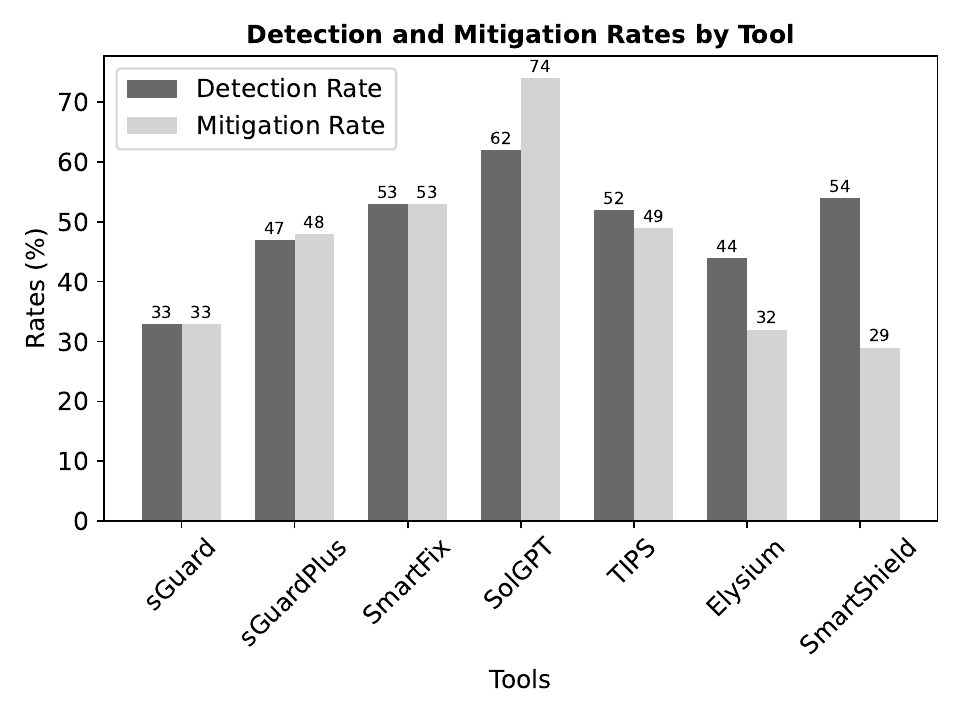}
    \caption{Detection Rate and Exploit Mitigation Rate for each APR tool. 
    Detection Rate represents the percentage of vulnerabilities with exploits (91) identified by each tool. Mitigation Rate reflects the percentage of exploits successfully mitigated out of the same total. Results are specific to vulnerabilities with exploits and, therefore,  differ from Table~\ref{tab:over_comparison_rq2}.}
    \label{fig:detection_mitgation}
\end{figure}

Smart contract vulnerability detection remains a significant challenge in the field.
Previous studies have highlighted limitations in detection capabilities~\cite{ChaliasosDeFiTools, VulnScanners}.
These detection limitations directly affect automated fixes: one cannot repair what one cannot detect.

Our empirical results quantify this dependency between detection accuracy and repair success.
Figure~\ref{fig:detection_mitgation} presents a comparison of each tool's detection rate (the percentage of exploitable vulnerabilities detected out of 91 total) and mitigation rate (the percentage of exploits mitigated out of 91).
For sGuard, SmartFix, and TIPS, the detection rate is very close to the mitigation rate, meaning high consistency.
For Elysium and SmartShield, the detection rate is higher, suggesting that repair is harder at the binary code level.

Both sGuard+ and SolGPT surprisingly mitigate more than what they detect.
sGuard+ mitigates one more exploit than the number of accurately detected vulnerabilities.
Manual analysis reveals this happens because the tool identifies and fixes a vulnerability of a different type than the labeled one.
The tool inadvertently mitigates the exploit targeting the labeled vulnerability by addressing this additional issue.
SolGPT exhibits a higher disparity between detection and mitigation rates: the mitigation rate (74\%) substantially exceeds the detection rate (62\%).
This is because the LLM changes the code beyond the given Slither warnings given in the prompt. 
In this case, the LLM leverages its training on large code repositories to regularize problematic code patterns directly, even when they are not explicitly prompted to do so.
We note that this behavior could result from training data leakage, yet we are not aware of systematic repositories with \dataset patches.
These results call for more research on LLM-based vulnerability detection for smart contracts.
\end{rblock}
\begin{rblock}
\subsection{Exploitability}
\input{tables/RQ3/not_exploitable}
A key insight of our study is that vulnerable does not necessarily imply exploitable, a conclusion consistent with prior work~\cite{vulnotexploit}.
In our analysis of the 143 contracts from the \dataset dataset, we find that 91/143 (63.6\%) can be successfully exploited using manually crafted exploits.
The remaining 52 contracts (36.4\%) could not be exploited, and the reasons for this are outlined in Table~\ref{tab:RQ3challenges}.
The majority of unexploitable cases (30 contracts) involve theoretical vulnerabilities. These issues are either gated behind owner-only access or lead to no adverse effect. In practice, these contracts are safe.
For 9 contracts, we found that the exploit entry point is missing. Although the vulnerability exists in theory, the contract lacks publicly accessible functions or interfaces that would allow an attacker to trigger the flaw.
In 8 contracts, Solidity version inconsistencies posed challenges.
1 contract uses an outdated version (v0.4.11), unsupported by modern tooling (e.g., Hardhat), making practical testing infeasible. The other 7 were mislabeled with incorrect compiler versions, leading to semantic differences that nullified the originally reported vulnerabilities.
3 contracts exceeded our analysis time budget of one hour, preventing us from confirming exploitability within reasonable bounds.
Finally, 2 contracts fall under unique cases. One is a honeypot, a deceptive contract designed to lure and trap would-be attackers. The other requires hash cracking as part of the exploit, which is computationally infeasible and beyond the scope of our analysis.



To sum up, we are the first to create an exploitable version of \dataset. From this process, the main takeaway is that it is hard to create a dataset of exploitable vulnerabilities. To our knowledge, there are very few such datasets,  Code4rena \cite{code4rena} is one notable example. Curating exploitable vulnerability datasets is of high value for the research community.
\end{rblock}

\begin{rblock}
\subsection{Bytecode vs. Source-level Repair}

Our results highlight a clear trade-off between bytecode-level and source-level APR tools. Bytecode-level tools like Elysium and SmartShield show the highest functional failure rates (29\% and 23\%, respectively), reflecting their struggle to preserve contract behavior, a critical concern for practitioner usage.

This limitation stems from bytecode's lack of semantic context. Without high-level constructs like variable names or control flow, these tools often apply low-level fixes that inadvertently alter intended behaviors. Moreover, patches generated at the bytecode level are difficult to interpret and to validate manually, further complicating correctness assurance.

Conversely, source-level tools such as SolGPT, SmartFix, and TIPS leverage semantic information to produce more accurate, human-readable patches. SolGPT, for instance, achieved the highest exploit mitigation rate (73\%) alongside strong functionality retention, demonstrating the advantages of semantically informed repair.

However, source-level tools require access to the contract’s source code, which is not always available in real-world deployments. In such scenarios, bytecode-level tools may be necessary due to their broader applicability, despite challenges in precision and validation.
\end{rblock}

\begin{rblock}
\subsection{Practical Recommendations for Future Research}
Our findings yield several practical insights that can inform future work:

\textbf{1. Relevance of Vulnerability Detection.} 
Although vulnerability detection is not the primary goal of automated program repair tools, it plays a critical role in their success.
We recommend the adoption of a standardized input format for vulnerabilities to enable interoperability between detection and repair components.
For example, the format that TIPS uses to specify detected vulnerabilities (see \autoref{lst:UP-TIPS}) is a good start towards a generic, machine readable format for specifying the vulnerabilities to be repaired.
This would allow users to plug in newer detection tools without modifying the repair pipeline.

\textbf{2. Patch Generation Strategies.}
Our study finds that the most effective approaches are generative AI (SolGPT) and mutation-based techniques (SmartFix), closely followed by template-based methods. 
We note that generative AI is a flexible generation engine. 
The model can be guided to emulate different repair paradigms, for example, few-shot prompting can emulate template-based repairs, while mutation-based strategies can be driven by prompting structured edits. 
We also foresee that future research can enhance generative AI repair with information from complementary components such as symbolic execution and fuzzing.

\textbf{3. Joint Patch Validation.}
In automated program repair for smart contracts, patch validation must account for both correctness and security. 
Proof-of-concept exploits offer a practical and effective way to assess whether a patch successfully mitigates a vulnerability.
This type of validation provides strong security guarantees. 
However, functional correctness is also essential and robust repair tools should ensure both security and correctness.
sb-heists provides APR research with a strong framework for validating patches for the sb-curated dataset. It is also a blueprint for future joint validation approaches for other benchmarks.

\end{rblock}

\section{Threats to Validity}\label{sec:validity}

\subsection{Internal Validity}
Our findings depend on the reliability of tool execution and the exploit validation checks.
In reproducing some APR tools, we encounter and resolve small errors. While done for enabling greater success, these fixes may affect the tool’s original functionality, potentially influencing the detection and patching processes. 
Additionally, our manual vulnerability analysis and exploit creation are subject to human judgment, which may lead to inconsistencies in determining whether a vulnerability is exploitable.
The functionality checks in our exploit evaluation may be incomplete, this could lead to misjudging the ability to mitigate exploits.
To mitigate this risk, all code and results have been discussed between the authors.

\subsection{External Validity}
The composition of the dataset directly influences the findings.
The \dataset dataset includes both real and artificial contracts, which may introduce biases since artificial contracts might not fully represent the complexity or nuances of real-world contracts. 

From an empirical perspective, the execution environment is also a threat for generalization.
The execution environment, using a local blockchain, may differ from the behavior of contracts on a live blockchain, potentially affecting the exploitability and repair effectiveness measurements.
To mitigate this threat, we use the industrial grade hardhat framework, used by the majority of the smart contract industry.

\begin{rblock}
    
\subsection{Representativeness of Considered Tools}
The APR tools we included in this comparative study represent 35\% (7 out of 20) of the academic tools we found.
This is because we did not manage to find or execute all existing tools, due to availability and scientific reproducibility issues.

We believe that those 7 tools we consider are representative of the state of the art for the following reasons. 
First, they were published across 3 out of the 4 years of active research in this area, hence capturing the evolution of the field. 
Second, they span fundamentally different repair strategies incl. template-based, mutation, and generative AI approaches. 
\end{rblock}

\section{Related Work }
\label{sec:related_work}

In this section, we discuss related work in the areas of  Smart Contract Repair Surveys and Smart Contract Vulnerabilities.

\subsection{Surveys}
There are notable surveys on smart contract repair.
Chu et al.~\cite{survey-data-detection-repair-2023} categorize repair tools into two types: off-chain tools, which address vulnerabilities before deployment, and on-chain tools, which repair deployed contracts.
While the survey highlights the advantages of each approach, it does not provide any discussion or experiment of actual vulnerability mitigation.
Qian et al.~\cite{qian2023empiricalreviewsmartcontract} compare repair methods and the types of vulnerabilities each tool targets.
Other very recent surveys have studied vulnerability repair and mitigation ~\cite{haouari2024vulnerabilitiessmartcontractsmitigation, VulnSCDetecMIt2024}, yet none of them has performed an empirical, quantitative comparison as we have done in this paper.

Our work is the first to empirically evaluate the effectiveness of these APR tools at generating patches (\autoref{subsec:rq2-methodology}), the first to validate functionality preservation and actual mitigation of vulnerabilities (\autoref{subsec:rq3-methodology}).

\subsection{Smart Contract Vulnerabilities}
An early study on vulnerabilities in smart contracts was conducted by Chen et al.~\cite{contractsdefect}, who identified 20 types of vulnerabilities.
Later, the work of Zou, et al.~\cite{challengesandopportunities}  surveyed 232 practitioners involved in smart contract development.
Among the many challenges raised, they highlighted the lack of proper tooling for ensuring security in smart contracts. 
These calls for action led to the development of various detection tools, which have since been the subject of numerous studies~\cite{ChaliasosDeFiTools,howeffective2020,2024surveydetectiontechniques}. 

Ghaleb et al.~\cite{howeffective2020} systematically evaluate smart contracts analysis tools by injecting known vulnerabilities. 
Chaliasos et. al.~\cite {ChaliasosDeFiTools} examine the effectiveness of vulnerability detection tools in terms of their potential to prevent attacks. 
Khan et al.~\cite{2024surveydetectiontechniques} propose a comprehensive literature review and classification of detection tools, which, to our knowledge, represents the most recent contribution in the rapidly evolving field of vulnerability detection.
From an empirical perspective the work of Durieux et al.~\cite{empirialevaldetectors2020durieux},  Di Angelo et al. ~\cite{diAngelo2024Empiricalevoldetectorsbytecode},  and Sendner et al.~\cite{VulnScanners} have assessed vulnerability detection tools by testing them on previously labeled datasets. 
These studies collectively emphasize the need for more accurate detection methods, as current tools tend to generate a high number of false positives and a concerning number of false negatives. 

To understand the high rate of false positives on vulnerability detection,
Perez et al.~\cite{vulnotexploit} propose to study the correlation between exploited and vulnerable contracts. 
They analyze on-chain smart contracts labeled as vulnerable by various detection tools and their actual exploitation status. 
The authors conclude that many contracts labeled as vulnerable are not exploitable.
Similarly, Sayeed et al. ~\cite{2020attacksandprotections} analyze attack vectors for seven vulnerabilities and highlight the inefficiency of widely used detection tools in preventing real-world exploits.
Jiao et al.~\cite{2024surveyethscsecurity-attackochdetection} provide hints of the attack vectors for eleven known vulnerabilities and categorize the different approaches for detection with their corresponding challenges. 
Finally, Zhang et al.~\cite{2023demystifyingexploitablebugs} demonstrate the value of detecting vulnerabilities from an exploit perspective, discovering 15 zero-day exploitable vulnerabilities in a study of 516 smart contracts.

Our work contributes to the much less researched area of smart contract repair. We are the first to demonstrate effectiveness through exploit mitigation, offering a sound evaluation of patch effectiveness.

\section{Conclusion}
\label{sec:conclusion}
This paper has introduced a novel framework for studying automated program repair of smart contracts using executable exploits. By creating exploits, by having sound functional checks, we provide the first concrete proof of vulnerability mitigation in the domain of smart contract security, offering the first reliable assessment of the state of the art of Automated Program Repair for smart contracts. Our methodology and reusable artifacts provide solid foundations for future research on automated repair for smart contracts.

\section*{Acknowledgment}
This work was partially supported by the WASP Program funded by Knut and Alice Wallenberg Foundation, and by the Swedish Foundation for Strategic Research (SSF). Some computation was enabled by resources provided by the National Academic Infrastructure for Supercomputing in Sweden (NAISS).


\balance
\bibliographystyle{plain}
\bibliography{main}



\end{document}

%% file: solidity/solidity-highlighting.tex
\usepackage{listings, xcolor}
\usepackage{relsize} 

\definecolor{verylightgray}{rgb}{.97,.97,.97}
\definecolor{lightgreen}{rgb}{0.839,0.961,0.839}

\lstdefinelanguage{Solidity}{
	keywords=[1]{anonymous, assembly, assert, balance, break, call, callcode, case, catch, class, constant, continue, constructor, contract, debugger, default, delegatecall, delete, do, else, emit, event, experimental, export, external, false, finally, for, function, gas, if, implements, import, in, indexed, instanceof, interface, internal, is, length, library, log0, log1, log2, log3, log4, memory, modifier, new, payable, pragma, private, protected, public, pure, push, require, return, returns, revert, selfdestruct, send, solidity, storage, struct, suicide, super, switch, then, this, throw, transfer, true, try, typeof, using, value, view, while, with, addmod, ecrecover, keccak256, mulmod, ripemd160, sha256, sha3}, 
	keywordstyle=[1]\color{blue}\bfseries,
	keywords=[2]{address, bool, byte, bytes, bytes1, bytes2, bytes3, bytes4, bytes5, bytes6, bytes7, bytes8, bytes9, bytes10, bytes11, bytes12, bytes13, bytes14, bytes15, bytes16, bytes17, bytes18, bytes19, bytes20, bytes21, bytes22, bytes23, bytes24, bytes25, bytes26, bytes27, bytes28, bytes29, bytes30, bytes31, bytes32, enum, int, int8, int16, int24, int32, int40, int48, int56, int64, int72, int80, int88, int96, int104, int112, int120, int128, int136, int144, int152, int160, int168, int176, int184, int192, int200, int208, int216, int224, int232, int240, int248, int256, mapping, string, uint, uint8, uint16, uint24, uint32, uint40, uint48, uint56, uint64, uint72, uint80, uint88, uint96, uint104, uint112, uint120, uint128, uint136, uint144, uint152, uint160, uint168, uint176, uint184, uint192, uint200, uint208, uint216, uint224, uint232, uint240, uint248, uint256, var, void, ether, finney, szabo, wei, days, hours, minutes, seconds, weeks, years},	
	keywordstyle=[2]\color{teal}\bfseries,
	keywords=[3]{block, blockhash, coinbase, difficulty, gaslimit, number, timestamp, msg, data, gas, sender, sig, value, now, tx, gasprice, origin},	
	keywordstyle=[3]\color{violet}\bfseries,
	identifierstyle=\color{black},
	sensitive=true,
	comment=[l]{//},
	morecomment=[s]{/*}{*/},
	commentstyle=\color{gray}\ttfamily,
	stringstyle=\color{red}\ttfamily,
	morestring=[b]',
	morestring=[b]"
}

\lstdefinestyle{diff}{
    escapechar=\%
}

\newcommand{\GHilight}{\makebox[0pt][l]{\color{lightgreen}\rule[-4pt]{1\linewidth}{10pt}}}
\newcommand{\RHilight}{\makebox[0pt][l]{\color{pink}\rule[-4pt]{1\linewidth}{10pt}}}

%% file: tables/Introduction/tools.tex
\begin{table*}[ht]
\centering
\caption{Overview of APR tools for smart contracts considered in this work. Each tool is described by its repair level, detection method (EXT: external, INT: internal, U-P: User-Provided), repair strategy, and patch validation approach (FV: Formal Verification, ME: Manual Evaluation, SA: Static Analysis). as reported in the original paper.}
\label{tab:ToolDescription}
\begin{tblr}{
  colspec ={l|l|p{0.38\linewidth}|l|l},
  row{4} = {gray!30!white},
  row{6} = {gray!30!white},
  row{8} = {gray!30!white},
  row{10} = {gray!30!white},
  row{12} = {gray!30!white},
  row{14} = {gray!30!white},
  row{16} = {gray!30!white},
  row{18} = {gray!30!white},
  row{20} = {gray!30!white},
  cell{2}{1} = {gray!30!white},
  cell{2}{2} = {r=17}{},
  cell{2}{3} = {gray!30!white},
  cell{2}{4} = {gray!30!white},
  cell{2}{5} = {gray!30!white},
  cell{19}{2} = {r=3}{},
  vline{1-3,6} = {1-2,19}{},
  vline{1-2,6} = {3-18,20-21}{},
  hline{1-2,19,22} = {-}{},
  hline{3-21} = {1,3-5}{},
}
Tool       & Level        & Vulnerability Detection  & Repair Strategy     & Patch Validation  \\ 

SCRepair~\cite{SCRepair}      (2020) & \textbf{source code} & EXT (Slither~\cite{slither}, Oyente~\cite{oyente}) & Mutation              & SA  \\ 
sGuard~\cite{sGuard}          (2021) & & INT & Template     & ME   \\  
Aroc~\cite{Aroc}              (2022) & & U-P & Template     & ME  \\ 
DeFinery~\cite{DeFinery}      (2022) & & U-P  & Mutation    & FV  \\ 
HCC~\cite{HCC}                (2022) & & INT & Template     & SA\&ME  \\  
ContractFix~\cite{ContractFix} (2023) & & EXT (Slither~\cite{slither},Securify~\cite{securify}, SmartCheck~\cite{smartcheck}) & Template     & SA\&ME  \\  
GPT\&BARD~\cite{gpt&bard}     (2023) & & EXT (Slither~\cite{slither}, Oyente~\cite{oyente}, Securify~\cite{securify}, HoneyBadger~\cite{honeybadger}, Osiris~\cite{osiris}, Mythril~\cite{mythril}) & Generative                      & SA   \\  
SmartFix~\cite{SmartFix}    (2023)  & & EXT (VeriSmart ~\cite{verismart})& Mutation              & FV\&ME  \\  
SmartRep~\cite{SmartRep}    (2023)  & & EXT & Supervised   & N/A  \\  
SolGPT~\cite{SolGPT}        (2023)  & & EXT (Slither~\cite{slither})& Generative                   & SA  \\  
TIPS~\cite{TIPS}            (2023)  & & EXT or U-P (Slither~\cite{slither}, Mythril~\cite{mythril},SmartEmbed~\cite{smartembed}) & Template           & SA\&ME  \\  
ReenRepair~\cite{ReenRepair} (2023) & & INT   & Template               & ME\&SA  \\  
ACFix~\cite{zhang2024acfix} (2024)  & & INT & Generative                   & SA\&ME \\  
RLRep~\cite{RLRep}          (2024)  & & EXT (Slither~\cite{slither}, Oyente~\cite{oyente}, Securify~\cite{securify}, Mythril~\cite{mythril}) & Supervised      & ME  \\  
sGuard+~\cite{sGuardPlus} (2024)  & & INT & Template    & ME \\  
vFix~\cite{fang2024vfix} (2024)             & & EXT (Securify~\cite{securify}, Slither~\cite{slither},  SmartCheck~\cite{smartcheck})& Template     & SA\&ME   \\  
ContractTinker ~\cite{contracttinker} (2024)&  & EXT (Slither~\cite{slither}) & Generative & SA\&ME \\ 

SmartShield~\cite{smartShield} (2020)& \textbf{bytecode}    & EXT (Securify~\cite{securify}, Osiris~\cite{osiris}, Mythril~\cite{mythril}) & Template & SA \\ 
EVMPatch~\cite{EVMPatch} (2021)& & EXT (Oyente~\cite{oyente}, Securify~\cite{securify}, Osiris~\cite{osiris}, ECF~\cite{ECF}, teEther~\cite{teEther},  Maian~\cite{maian}, Sereum~\cite{sereum}) & Template               & SA  \\ 
Elysium~\cite{Elysium} (2022) & & EXT (Oyente~\cite{oyente}, Osiris~\cite{osiris}, Mythril~\cite{mythril}) & Template     & SA  \\ 
\end{tblr}
\end{table*}

%% file: solidity/contract_reentrancy.tex
\begin{minipage}{\linewidth}
\begin{lstlisting}[language=solidity,label={lst:dataset-example-reentrancy},caption={Example of a \dataset smart contract with a reentrancy vulnerability.}]
contract Reentrancy {
  mapping (address => uint) userBalance;
  ...
  function withdrawBalance() {
    // <yes> <report> REENTRANCY
    if(!(msg.sender.call.value(userBalance[msg.sender])())){
      throw;
    }
    userBalance[msg.sender] = 0;
  }
}
\end{lstlisting}
\end{minipage}

%% file: tables/RQ1/results.tex
\begin{table}
\centering
\caption{Results for RQ1: Tools labeled according to their availability, and whether Installation and Execution was a success (~\cmark) or failure (~\xmark)}
\label{tab:RQ1results}
\begin{tblr}{
  cells = {c},
  row{even} = {gray!30!white},
  hline{1-2} = {-}{},
}
\textbf{Tool}      & \textbf{Availability}                                                           & \textbf{Installability}  & \textbf{Executability} \\ 
SCRepair           & \href{https://github.com/xiaoly8/SCRepair?tab=readme-ov-file}{GitHub}           & \xmark           & -                \\ \hline
sGuard             & \href{https://github.com/duytai/sGuard}{GitHub}                                 & \cmark           & \cmark           \\ \hline
DeFinery           & \href{https://github.com/palinatolmach/DeFinery}{GitHub}                        & \xmark           & -                \\ \hline
HCC                & Not Found                                                                       & -           & -                \\ \hline
ContractFix        & \href{https://github.com/research1132/ContractFix}{GitHub}                      & \xmark           & -                \\ \hline
GPT\&BARD          & Not Found                                                                       & -           & -                \\ \hline
SmartFix           & \href{https://zenodo.org/records/8256377}{Zenodo}                               & \cmark           & \cmark           \\ \hline

\SetCell[r=2]{c} SmartRep & \SetCell[r=1]{c}{\href{https://github.com/smartrep}{GitHub}} & \SetCell[r=1]{c} \cmark &
\makecell{
    \begin{tabular}{@{}c|c@{}}
    \textbf{Training} & \textbf{Repair} \\
    \hline
    \xmark & \xmark
    \end{tabular}
} \\ \hline
SolGPT             & \href{https://github.com/enaples/SolGPT}{GitHub}                                & \cmark           & \cmark           \\ \hline
TIPS               & \href{https://github.com/CVbluecat/TIPS}{GitHub}                                & \cmark           & \cmark           \\ \hline
ACFix              & \href{https://anonymous.4open.science/r/Repair-Access-Control-C-B377/README.md}{GitHub} & \cmark   & \xmark           \\ \hline
\SetCell[r=1]{c} RLRep & \SetCell[r=1]{c}{\href{https://github.com/rlrep}{GitHub}} & \SetCell[r=1]{c} \cmark &
\makecell{
    \begin{tabular}{@{}c|c@{}}
    \textbf{Training} & \textbf{Repair} \\
    \hline
    \xmark & \xmark
    \end{tabular}
} \\ \hline
sGuard+         & \href{https://zenodo.org/records/8249340}{Zenodo}                               & \cmark           & \cmark           \\ \hline
vFix               & \href{https://github.com/vfixresearch/vFix/tree/main}{GitHub}                   & \cmark           & \xmark                \\ \hline
ContractTinker     & \href{https://github.com/CheWang09/LLM4SMAPR/}{GitHub}                          & \cmark           & \xmark                \\ \hline
SmartShield        & \href{https://smartshield.code-analysis.org/}{Site}                             & \cmark           & \cmark           \\ \hline
EVMPatch           & Non Available                                                                   & -           & -                \\ \hline
Elysium            & \href{https://github.com/christoftorres/Elysium}{GitHub}                        & \cmark           & \cmark           \\ \hline
ReenRepair         & Non Available                                                                   & -           & -                \\ \hline
\end{tblr}
\end{table}

%% file: tables/RQ2/results.tex
\begin{table*}
\centering
\caption{RQ2: Comparison of patches generated for each tool based on detection, compilation, modification, and vulnerability repair. \textbf{C} represents the total number of vulnerable contracts with at least one patch in that category,   and \textbf{P} represents the number of patches in that category.}
\label{tab:over_comparison_rq2}
\begin{tblr}{
  row{2} = {c},
  row{5} = {gray!30!white},
  row{9} = {gray!30!white},
  column{3} = {c},
  cell{1}{1} = {r=2}{},
  cell{1}{2} = {r=2}{},
  cell{1}{3} = {r=2}{},
  cell{1}{4} = {c=2}{c},
  cell{1}{6} = {c=2}{c},
  cell{1}{8} = {c=2}{c},
  cell{1}{10} = {c=2}{c},
  cell{3}{1} = {r=4}{},
  cell{3}{2} = {gray!30!white},
  cell{3}{3} = {gray!30!white},
  cell{3}{4} = {gray!30!white,c},
  cell{3}{5} = {gray!30!white,c},
  cell{3}{6} = {gray!30!white,c},
  cell{3}{7} = {gray!30!white,c},
  cell{3}{8} = {gray!30!white,c},
  cell{3}{9} = {gray!30!white,c},
  cell{3}{10} = {gray!30!white,c},
  cell{3}{11} = {gray!30!white,c},
  cell{4}{4} = {c},
  cell{4}{5} = {c},
  cell{4}{6} = {c},
  cell{4}{7} = {c},
  cell{4}{8} = {c},
  cell{4}{9} = {c},
  cell{4}{10} = {c},
  cell{4}{11} = {c},
  cell{5}{4} = {c},
  cell{5}{5} = {c},
  cell{5}{6} = {c},
  cell{5}{7} = {c},
  cell{5}{8} = {c},
  cell{5}{9} = {c},
  cell{5}{10} = {c},
  cell{5}{11} = {c},
  cell{6}{4} = {c},
  cell{6}{5} = {c},
  cell{6}{6} = {c},
  cell{6}{7} = {c},
  cell{6}{8} = {c},
  cell{6}{9} = {c},
  cell{6}{10} = {c},
  cell{6}{11} = {c},
  cell{7}{2} = {gray!30!white},
  cell{7}{3} = {gray!30!white},
  cell{7}{4} = {gray!30!white,c},
  cell{7}{5} = {gray!30!white,c},
  cell{7}{6} = {gray!30!white,c},
  cell{7}{7} = {gray!30!white,c},
  cell{7}{8} = {gray!30!white,c},
  cell{7}{9} = {gray!30!white,c},
  cell{7}{10} = {gray!30!white,c},
  cell{7}{11} = {gray!30!white,c},
  cell{8}{1} = {r=2}{},
  cell{8}{4} = {c},
  cell{8}{5} = {c},
  cell{8}{6} = {c},
  cell{8}{7} = {c},
  cell{8}{8} = {c},
  cell{8}{9} = {c},
  cell{8}{10} = {c},
  cell{8}{11} = {c},
  cell{9}{4} = {c},
  cell{9}{5} = {c},
  cell{9}{6} = {c},
  cell{9}{7} = {c},
  cell{9}{8} = {c},
  cell{9}{9} = {c},
  cell{9}{10} = {c},
  cell{9}{11} = {c},
  cell{10}{4} = {c=2}{c},
  cell{10}{6} = {c=2}{c},
  cell{10}{8} = {c=2}{c},
  cell{10}{10} = {c=2}{c},
  vlines, 
  hline{1,3,8,10-11} = {-}{},
}
\textbf{Level}           & \textbf{Tool} & \textbf{detected C} & \textbf{generated} &     & \textbf{compilable} &     & \textbf{different}   &     & \textbf{consistent}    &     \\
                         &               &                     & \textbf{C}         & P   & \textbf{C}          & P   & \textbf{C}           & P   & \textbf{C}             & P   \\
\textbf{source code}     & sGuard        & 35                  & \textbf{109}       & 109 & \textbf{108~(99\%)} & 108 & \textbf{62~(\revision{57\%})}   & 62  & \textbf{2/35~(\revision{6\%})}    & 2   \\
                         & sGuard+    & 70                  & \textbf{81}        & 81  & \textbf{81~(100\%)} & 81  & \textbf{81~(100\%)}  & 81  & \textbf{70/70~(100\%)} & 70  \\
                         & SmartFix      & 51                  & \textbf{86}        & 86  & \textbf{86~(100\%)} & 86  & \textbf{86~(100\%)}  & 86  & \textbf{48/51~(94\%)}  & 48  \\
                         & SolGPT        & 97                  & \textbf{139}       & 552 & \textbf{138~(99\%)} & 527 & \textbf{139~(100\%)} & 552 & \textbf{88/97~(\revision{91\%})}  & 332 \\
                         & TIPS          & 82                  & \textbf{140}       & 242 & \textbf{138~(\revision{99\%})} & 234 & \textbf{140~(100\%)} & 242 & \textbf{81/82~(\revision{99\%})}  & 129 \\
\textbf{bytecode}        & Elysium       & 52                  & \textbf{121}       & 121 & \textbf{n/a}        & n/a & \textbf{95~(\revision{79\%})}   & 95  & \textbf{52/52~(100\%)} & 52  \\
                         & SmartShield   & 59                  & \textbf{135}       & 135 & \textbf{n/a}        & n/a & \textbf{135~(100\%)} & 135 & \textbf{38/59~(64\%)}  & 38  \\
\textbf{Total Contracts} & \textbf{143}  & \textbf{119}        & \textbf{142}       &     & \textbf{142}        &     & \textbf{142}         &     & \textbf{115}           &     
\end{tblr}
\end{table*}

%% file: solidity/tips_patch_diff.tex
\begin{minipage}{\linewidth}
\begin{lstlisting}[language=Solidity,style=diff,label={lst:invalid_patch},caption={Reentrancy vulnerability patched by TIPS~\cite{TIPS}.}]
mapping (address => uint) userBalance;
...
function withdrawBalance() {
%\GHilight%+  userBalance[msg.sender] = 0;
  if(!(msg.sender.call.value(userBalance[msg.sender])())){
      throw;
  }
%\RHilight%-  userBalance[msg.sender] = 0;
}
\end{lstlisting}
\end{minipage}

%% file: solidity/smartfix_patch_diff.tex
\begin{minipage}{\linewidth}
\begin{lstlisting}[language=Solidity,style=diff,label={lst:effective_patch},caption={Reentrancy vulnerability patched by SmartFix~\cite{SmartFix}.}]
mapping (address => uint) userBalance;
...
function withdrawBalance() {
%\GHilight%+  uint256 tmp_1 = userBalance[msg.sender];
%\GHilight%+  userBalance[msg.sender] = 0;
  if(!(msg.sender.call.value(tmp_1)())){
      throw;
  }
%\RHilight%-  userBalance[msg.sender] = 0;
}
\end{lstlisting}
\end{minipage}

%% file: tables/RQ3/mitigated_exploits_by_tool.tex
\begin{table}
\centering
\caption{RQ3: Number of patched contracts that failed functional checks and the total and unique exploits mitigated by each tool out of 91 exploits.}
\label{tab:RQ3results}
\begin{tblr}{
  cells = {c},
  row{3} = {gray!30!white},
  row{5} = {gray!30!white},
  row{7} = {gray!30!white},
  row{9} = {gray!30!white},
  cell{1}{1} = {r=2}{},
  cell{1}{2} = {r=2}{},
  cell{1}{3} = {c=2}{},
  hline{1,3,10} = {-}{},
  hline{2} = {3-4}{},
}
\textbf{Tool} & {\textbf{Failed}\\\textbf{Functional Checks}} & \textbf{Mitigated Exploits} &                 \\
              &                                           & \textbf{Total}              & \textbf{Unique} \\
sGuard & 1 (1\%) & 30 (\revision{33\%}) & 0 \\
sGuardP+ & 0 (0\%) & 44 (48\%) & 0 \\
SmartFix & 4 (4\%) & 48 (\revision{53\%}) & 1 \\
SolGPT & 7 (\revision{8\%}) & 67 (\revision{74\%}) & 4 \\
TIPS & 1 (1\%) & 45 (49\%) & 4 \\
Elysium & 26 (\revision{29\%}) & 29 (\revision{32\%}) & 1 \\
SmartShield & 21 (23\%) & 26 (\revision{29\%}) & 0 \\
\end{tblr}
\end{table}

%% file: tables/RQ4/results.tex

\begin{table*}
\centering
\caption{\revision{RQ4: Mitigation effectiveness per vulnerability type, tool and contract in dataset \dataset}}
\label{tab:rq4_results}
\begin{tblr}{
  width = \textwidth,
  colspec = {X[c,m,wd=2cm] X[c,m,wd=1cm] *{6}{Q[c,m,h=1.3cm,wd=0.9cm]} Q[c,m,h=1.3cm,wd=1.2cm] Q[c,m,wd= 1.3cm] Q[c,m,wd=1.7cm]},
  row{even} = {gray!30!white},
  row{1} = {bg=gray!30, font=\bfseries},
  row{2} = {bg=gray!30, font=\bfseries},
  row{Z} = {bg=gray!30, font=\bfseries},
  hline{13} = {1-13}{solid},
  vline{2,3, 10} = {1-13}{solid}, 
  cell{1}{3} = {c=7}{c},
  cell{1}{10} = {c=2}{c},
  cell{1}{1} = {r=2}{c},
  cell{1}{2} = {r=2}{c},
}
Vulnerability Type & \# Exploits & \SetCell[c=7]{c} Tools & & & & & & & \SetCell[c=2]{c} Aggregation \\
& & \rotatebox{30}{sGuard} & \rotatebox{30}{sGuard+} & \rotatebox{30}{SmartFix} & \rotatebox{30}{SolGPT} & \rotatebox{30}{TIPS} & \rotatebox{30}{Elysium} & \rotatebox{30}{SmartShield} & \# Mitigation & \% Effectiveness \\
Unchecked Low-Level Calls & 20 & n/a & 5 & n/a & 15 & 18 & 8 & 9 & 19 & 95\% \\  
Reentrancy & 26 & 25 & 25 & 26 & 25 & 25 & 0 & 4 & 26 & 100\% \\
Access Control & 16 & 3 & 4 & 10 & 11 & 2 & 8 & n/a (2) & 13 & 81\% \\
Arithmetic & 13 & 2 & 10 & 12 & 9 & n/a & 7 & 9 & 13 & 100\% \\
Bad Randomness & 4 & n/a & n/a & n/a & 3 & n/a & n/a (3) & n/a & 4 & 100\% \\
Denial of Service & 4 & n/a & n/a & n/a & n/a & n/a & n/a & n/a & 0 & 0\% \\
Time Manipulation & 3 & n/a & n/a & n/a & 2 & n/a & n/a (2) & n/a(1) & 2 & 67\% \\
Front Running & 3 & n/a & n/a & n/a & n/a & n/a & n/a (1) & n/a (1) & 1 & 33\% \\
Miscellaneous & 2 & n/a & n/a & n/a & 2 & n/a & n/a & n/a & 2 & 100\% \\
\textbf{Total} & \textbf{91} & 30 & 44 & 48 & 67 & 45 & 29 & 26 & \textbf{80} & \textbf{88\%} \\
    
\end{tblr}
\end{table*}

%% file: tables/RQ3/not_exploitable.tex
\begin{table}
\centering
\caption{Distribution of Unexploitable SmartBugs-Curated Contracts by Challenge Type.}
\label{tab:RQ3challenges}
\begin{tblr}{
  cells={c},
  row{even} = {gray!30!white},
  hline{1-2,9} = {-}{},
}
\textbf{Challenge Type}               & \textbf{Number of Contracts} \\
Theoretical Problem               & 30                 \\
Missing Exploit Entry Point               & 9                  \\
Solidity Version   & 8                  \\
Exceeded Analysis Time Limit           & 3                        \\
Honeypot and Hash Cracking             & 2                 \\
\end{tblr}
\end{table}